\documentclass[twocolumn,showpacs,%
  nofootinbib,aps,superscriptaddress,%
  eqsecnum,prd,notitlepage,showkeys,10pt]{revtex4-1}

\usepackage{amssymb}
\usepackage{amsmath}
\usepackage{graphicx}
\usepackage{dcolumn}
\usepackage{parskip}
\usepackage{graphicx}
\usepackage{enumitem}
\usepackage{makecell}
\usepackage{comment}
\usepackage{algorithm}
\usepackage{algpseudocode}
\usepackage{float}
\usepackage{subcaption}
\usepackage{mathtools}
\usepackage[hidelinks]{hyperref}
\usepackage[font=small,labelfont=bf]{caption}

\usepackage{etoolbox}
\AtBeginEnvironment{thebibliography}{\hypersetup{hidelinks, colorlinks=false, pdfborder={0 0 0}}}

\newcommand{\bt}{\boldsymbol{\theta}}
\newcommand{\bp}{\boldsymbol{p}}

\usepackage[utf8]{inputenc}  % ensures LaTeX reads the file as UTF-8
\usepackage[T1]{fontenc}     % ensures proper font encoding
\usepackage{textcomp}  

\hypersetup{
  colorlinks   = true, %Colours links instead of ugly boxes
  urlcolor     = blue, %Colour for external hyperlinks
  linkcolor    = blue, %Colour of internal links
  citecolor   = red %Colour of citations
}

\begin{document}

% Temporarily fix spacing for title/authors
\newlength{\oldparskip}
\setlength{\oldparskip}{\parskip}
\setlength{\parskip}{0pt}

\title{A Binary Optimisation Algorithm for Near-Term Photonic Quantum Processors}

\author{Alexander Makarovskiy}
\email{amakarovskiy@orcacomputing.com}
\affiliation{ORCA Computing}

\author{Mateusz Slysz}
\affiliation{Poznań Supercomputing and Networking Center, IBCH PAS}
\affiliation{Poznań University of Technology}

\author{Łukasz Grodzki}
\affiliation{Poznań Supercomputing and Networking Center, IBCH PAS}
% \affiliation{Poznań University of Technology}

\author{Dawid Siera}
\affiliation{Poznań Supercomputing and Networking Center, IBCH PAS}
% \affiliation{Poznań University of Technology}

\author{Thorin Farnsworth}
\affiliation{ORCA Computing}

\author{William R. Clements}
\affiliation{ORCA Computing}

\author{Piotr Rydlichowski}
\affiliation{Poznań Supercomputing and Networking Center, IBCH PAS}

\author{Krzysztof Kurowski}
\affiliation{Poznań Supercomputing and Networking Center, IBCH PAS}
% \affiliation{Poznań University of Technology}

\begin{abstract}
Binary optimisation tasks are ubiquitous in areas ranging from logistics to cryptography. The exponential complexity of such problems means that the performance of traditional computational methods decreases rapidly with increasing problem sizes. Here, we propose a new algorithm for binary optimisation, the Bosonic Binary Solver, designed for near-term photonic quantum processors. This variational algorithm uses samples from a quantum optical circuit, which are post-processed using trainable classical bit-flip probabilities, to propose candidate solutions. A gradient-based training loop finds progressively better solutions until convergence. We perform ablation tests that validate the structure of the algorithm. We then evaluate its performance on an illustrative range of binary optimisation problems, using both simulators and real hardware, and perform comparisons to classical algorithms. We find that this algorithm produces high-quality solutions to these problems. As such, this algorithm is a promising method for leveraging the scalable nature of photonic quantum processors to solve large-scale real-world optimisation problems.
\end{abstract}

\maketitle

% Restore normal spacing for the main document
\setlength{\parskip}{\oldparskip}

\section{Introduction}
One of the most promising applications of quantum computing, especially in the NISQ (Noisy Intermediate-Scale Quantum) era of quantum processors,  is discrete optimisation \cite{hybridsolverreview,hybridoptimisationreview,variationalquantumalgorithms}. Particularly, the Quantum Approximate optimisation Algorithm (QAOA), Variational Quantum Eigensolvers (VQEs) and quantum annealing protocols have yielded promising results with currently available hardware \cite{QAOA_hardware,VQE_Hardware,dwave_optimisation_review}. Both QAOA and VQE adopt a variational approach in which a parameterised quantum circuit is used to approximate the optimal solution through variational minimization. However, algorithmic development of these approaches has focused on universal gate-based architectures. On the other hand, quantum annealing uses special-purpose processors that initialise a quantum system in the ground state of a simple Hamiltonian and slowly evolves it towards a problem Hamiltonian that encodes the cost function \cite{kurowski_2020}.

Photonics has been identified as a promising path towards scalable quantum computing architectures \cite{Photonic_QC_review, knill_scheme_2001,O_Brien_2009}. Several large-scale near-term photonic quantum computing systems involving hundreds of photons and modes have been demonstrated to date \cite{USTC,borealis, USTC_2}.
The scalability of photonics is underpinned by some crucial advantages compared to other quantum architectures such as the absence of decoherence, fast calculation speeds, native networking, and room-temperature operation. These attributes would be highly desirable for running practical optimisation problems.
However, current photonic systems are limited to non-universal quantum computing paradigms such as Boson Sampling, which lacks the full set of operations needed to implement algorithms such as QAOA or VQE. As such, quantum optimisation algorithms that have been developed for other platforms are currently not suitable for near-term photonic quantum processors.

In this work, we present a hybrid classical-quantum optimisation algorithm that utilises near-term photonic architectures. Similarly to VQE, this Bosonic Binary Solver (BBS) algorithm uses a quantum optical circuit as an ansatz that is optimised within a classical training loop to produce increasingly better solutions to a target problem. Classical post-processing allows this algorithm to both reach the entire solution space and find better solutions. Importantly, BBS is not restricted to solving problems posed in a quadratic unconstrained binary optimisation (QUBO) format as is the case for quantum annealers, and has no quantum hardware restrictions on the classes of cost function it can be applied to as is the case for QAOA. This allows optimisation problem implementations with lower overheads in the length of the encoding. We show using both simulated and real quantum hardware that this algorithm finds good solutions on a range of discrete optimisation tasks such as Knapsack optimisation, Tactical Deconfliction and the travelling salesperson problem (TSP). This work thus paves the way for leveraging scalable near-term photonic quantum computing platforms for solving large-scale real-world optimisation problems.

\section{Background}
In this section, we introduce quantum optical circuits and their implementation using near-term quantum processors. We also discuss prior work on their use for optimisation.

In a photonic quantum processor, $n$ identical photons are sent into a quantum optical circuit of $m$ modes. The circuit consists of a set of optical beamsplitters and phase shifters, with reflectivities typically determined by a programmable parameter. Photon interference within this circuit creates an entangled state. A measurement is then performed to collapse this state and return the output locations of the photons. Sampling from the output distribution of a quantum optical circuit is known as boson sampling and is computationally hard to simulate classically \cite{AaronsonArkhipov}.

\subsection{Time-bin photonic quantum processors}

\begin{figure}[h]
    \centering
    \includegraphics[width=0.45\textwidth]{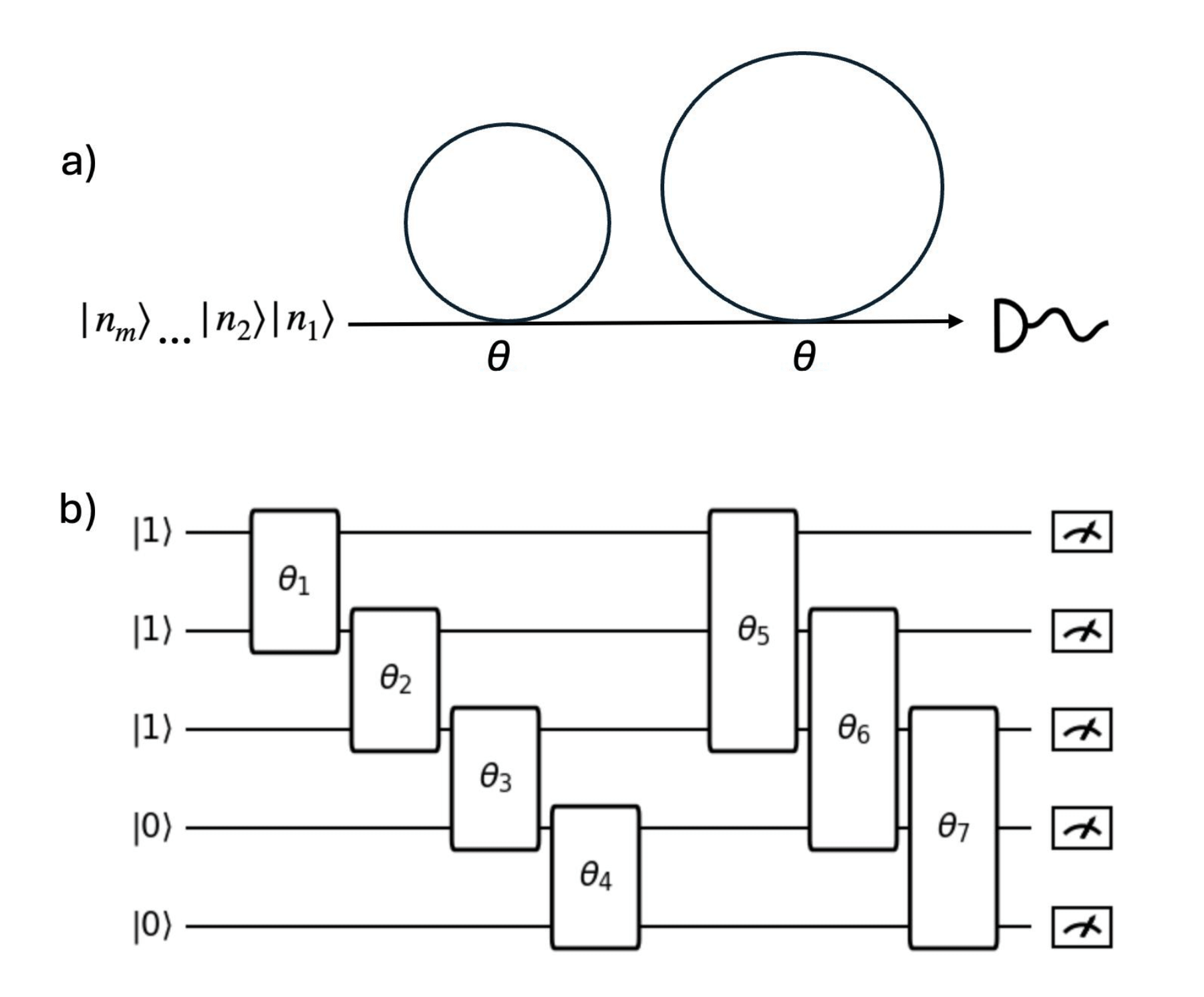}
    \caption{\textbf{a)} In this work, our experiments focus on quantum photonic processors that use a time-bin architecture. In such an architecture, single photons are sequentially sent into a network of optical delay lines with programmable coupling coefficients. This creates an entangled state between different time-bins which can be measured by a photon number resolving detector. \textbf{b)} Photonic circuit implemented by the time-bin architecture shown above, where the length of the first delay line is such that sequential photons interfere with each other, and the second delay line is twice as long. Each operation is a beamsplitter between two modes with a programmable coupling parameter. This example interferes 3 photons in 5 modes.}
    \label{fig:tbi}
\end{figure}

In this work, we focus on quantum optical circuits that use a time-bin architecture \cite{LoopBasedBosonSampling, He_2017}. While several physical realisations of quantum optical circuits exist—including spatial-mode and free-space interferometers—architectures based on optical delay lines, such as loop-based setups, have been gaining traction \cite{borealis, GBS_deshpande,novak2024boundaries}. This is largely due to their inherent scalability: they allow multiple time-bin modes to be processed using a single photon source and detector, greatly reducing the experimental overhead. These architectures are not only well-studied in the literature but are also commercially available, making them a practical choice for algorithm development. A diagram of a time-bin photonic system and the corresponding circuit diagram is shown in figure \ref{fig:tbi}.

The specific type of time-bin quantum optical circuit we will consider is a "power-law" architecture, consisting of three sequential delay lines that interfere photons separated by 1, 3 and 9 time-bins respectively. This is motivated by prior work on these architectures in \cite{GBS_deshpande,novak2024boundaries}, showing that they support long-range entanglement and are likely still hard to simulate classically despite a low physical component count. Moreover, though our experiments on real hardware use only a single delay line within the ORCA Computing PT-1, future PT Series architectures are expected to implement power-law architectures. Our results with this architecture in simulation can thus help predict the performance of this algorithm running on real hardware in the near-term.

\subsection{Binary Optimisation with Photonic Quantum Processors}

We review work done previously demonstrating how we can leverage quantum optical circuits to find solutions to binary optimisation tasks. We consider binary optimisation tasks where, for the set of all binary strings of a given length $m$, we define a \emph{cost function} $C: \{0,1\}^m \rightarrow \mathbb{R}$ and we wish to find the binary string with the lowest cost function value:
$$\boldsymbol{x} = \text{argmin} \ C(\boldsymbol{x}).$$
This class of binary optimisation problems encompasses a wide variety of computationally hard problems, including instances of Max-Cut, subset sum, knapsack, and various constraint satisfaction problems. Many of these are NP-hard and appear frequently in fields such as operations research, machine learning, and network science. 

The general framework we consider maps each output sample from the circuit to a candidate solution. Since sampling from the circuit is a computationally hard task, this framework allows us to sample from the space of possible solutions to the problem in a non-classical and potentially advantageous way. However, due to the non-universal nature of quantum optical circuits it is not obvious how best to perform this mapping in a way that can cover the entire solution space.

A method of mapping boson sampling statistics to candidate solutions for a binary optimisation task was first proposed in \cite{BradlerWallner}. It was proven that, with 4 different configurations of $m$ mode boson samplers with input single-photon states and parity measurements, we can measure any possible bit string of length $m$. A binary optimisation algorithm must be able to generate any possible bit string as a candidate solution, otherwise it may not be able to produce the optimal solution. The algorithm, denoted VBS (Variational Bosonic Solver), necessitated using single-photon input states of up to $m$ photons, and parity photon measurements. Most near-term quantum optical circuits work in regimes where the number of detected photons is significantly less than the total number of modes \cite{USTC,borealis}, such that detecting $m$ photons in $m$ modes is a hardware challenge. This made VBS impractical as a method of performing optimisation with real near-term devices.

Algorithms mapping statistics from a photonic processor with Gaussian input states to candidate solutions for QUBO problems were later studied in \cite{GBS_Optimisation, Training_GBS}. In both cases, threshold detectors were used without post-processing, allowing for complete coverage of the solution space due to the Gaussian input states being able to achieve any possible detection pattern. Post-processing would however be required to reach the entire solution space when single-photon input states are used. 

Here, we introduce a new algorithm, the \emph{Bosonic Binary Solver} (BBS) that overcomes these challenges and is suitable for experimentally realisable near-term photonic systems. BBS uses classical post-processing bit-flips to allow the algorithm to generate every possible bit string with a single $m$ mode boson sampler configuration with an arbitrary input state and photon click detectors. An early version of this algorithm was investigated in \cite{Slysz,Benchmarking_PT1_Paper}. In this work, we go beyond this by adding a novel, more efficient, gradient calculation method for circuit training, and a new approach to sample tracking. We also rigorously outline the framework of the algorithm, perform ablation tests, and benchmark performance.

\section{Theory}
\subsection{The Bosonic Binary Solver}

\begin{figure}[h]
    \centering
    \includegraphics[width=0.45\textwidth]{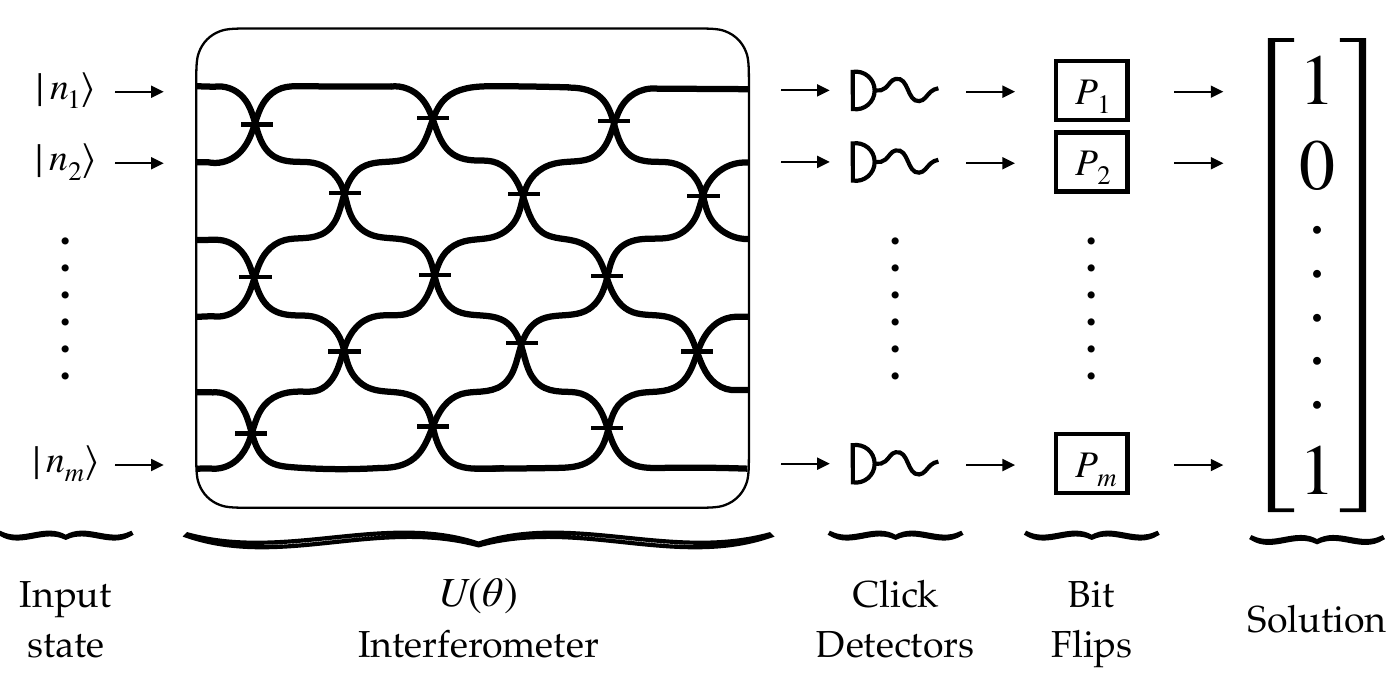}
    \caption{Schematic diagram of the BBS algorithm architecture. A single run of the quantum optical circuit with click detectors produces a binary measurement pattern. In a classical post-processing step, trainable probabilistic bit-flips are applied, returning a new bit string. A collection of bit strings produced using this process is used to estimate the average cost of generated solutions, and the parameters of the quantum circuit and the bit-flips are updated accordingly. This process continues with the circuit iteratively improving the average quality of solutions generated.}
    \label{fig:bbs_diagram}
\end{figure}

We set up the problem as follows. We consider a size $m$ binary optimisation task and an $m$-mode interferometer consisting of beamsplitters and phaseshifters parametrised by the variables $\boldsymbol{\theta}$. We denote the unitary matrix applied by the interferometer by $U(\bt)$. 

\begin{algorithm}[H]
\caption{Bosonic Binary Solver}\label{alg:bbs}
\begin{algorithmic}[1]
\Statex \hspace*{-\algorithmicindent} \textbf{Inputs:} Cost function $C$, input state $\lvert \mathbf{n} \rangle$, parametrised interferometer scheme $U(\bt)$, update number $N$, sample number $S$
\Statex \hspace*{-\algorithmicindent} \textbf{Outputs:} Lowest cost value found and corresponding solution $C_{\text{alg}},  \mathbf{x}_{\text{alg}}$
\State $\theta_i \leftarrow \text{ uniform random }(0, 2\pi) \quad \forall i$
\State $p_i \leftarrow 1/2 \quad \forall i$
\For{$j \leftarrow 1$ \text{to} $N$}
  \For{$k \leftarrow 1$ \text{to} $S$}
    \State Evolve $\lvert \mathbf{n} \rangle$ by $U(\bt)$
    \State{$\mathbf{x} \leftarrow$ threshold measurements}
    \State Store $\mathbf{x}$
    \State{$\mathbf{x} \leftarrow$ apply $\mathbf{p}$ bit-flips}
    \State Calculate cost function value $C(\mathbf{x})$
   \EndFor
    \For{$\theta_i \in \bt$}
      \State{$\bt_{up} = (\theta_1, ..., \theta_i + \delta, ..., \theta_m)$}
      \State{$\bt_{down} = (\theta_1, ..., \theta_i - \delta, ..., \theta_m)$}
      \For{$l \leftarrow 1$ \text{to} $S$} 
        \State{Evolve $\lvert \mathbf{n} \rangle$ by $U(\bt_{up})$ and $U(\bt_{down})$}
        \State{$\mathbf{x}_{i,up}, \mathbf{x}_{i,down} \leftarrow$ threshold measurements}
        \State{$\mathbf{x}_{i,up}, \mathbf{x}_{i,down} \leftarrow$ apply $\mathbf{p}$ bit-flips}
        \State{Calculate $C(\mathbf{x}_{i,up}), C(\mathbf{x}_{i,down})$}
      \EndFor
    \State Calculate $\frac{\partial C}{\partial \theta_i}$ using $C(\mathbf{x}_{i,up}), C(\mathbf{x}_{i,down})$ values
    \EndFor
    \For{$p_i \in \mathbf{p}$}
      \State{$\bp_{up} = (p_1, ..., p_i = 1, ..., p_m)$}
      \State{$\bp_{down} = (p_1, ..., p_i =0, ..., p_m)$}
      \For{$l \leftarrow 1$ \text{to} $S$}
        \State{Load previously measured string $\mathbf{x}$}
        \State{$\mathbf{x}_{i,up} \leftarrow$ apply $\bp_{up}$ bit-flips to $\mathbf{x}$}
        \State{$\mathbf{x}_{i,up} \leftarrow$ apply $\bp_{down}$ bit-flips to $\mathbf{x}$}
        \State{Calculate $C(\mathbf{x}_{i,up}), C(\mathbf{x}_{i,down})$}
      \EndFor
    \State Calculate $\frac{\partial C}{\partial p_i}$ using $C(\mathbf{x}_{i,up}), C(\mathbf{x}_{i,down})$ values
    \EndFor
    \State{$\bt,\bp \leftarrow$ update w.r.t. gradients according to optimiser}
\EndFor
\Statex \hspace*{-\algorithmicindent} \textbf{Return} Best found $\mathbf{x}_{alg}, C_{alg}$ out of all calculated
\end{algorithmic}
\end{algorithm}

Our Bosonic Binary Solver algorithm is described in Algorithm \ref{alg:bbs}. In summary, BBS injects a fixed input state of photons into an $m$-mode circuit with randomly initialised parameters. We draw samples from the evolved state with threshold detector measurements. Alternatively, photon number resolving measurements can be post-processed to the same effect. This set of bit strings is then post-processed using a set of trainable per-mode probabilistic bit-flip parameters. We then determine the average cost function value over these post-processed bit strings. Subsequently, the gradients of the average sample cost function with respect to both the circuit parameters and the bit-flip parameters are estimated through sampling (discussed in detail later), and these parameters are updated. This process repeats for a given number of update steps, after which we return the best solution found by the algorithm. The best solution is pulled from all bit strings for which the cost function has been calculated over the course of training, including those generated for gradient calculation.

This algorithm introduces two novel components compared to the original algorithm proposed in \cite{BradlerWallner}. First, where the original proposal in \cite{BradlerWallner} requires detecting $m$ photons in $m$ modes, BBS can explore the entire solution space with an arbitrary input state. This makes BBS significantly easier to implement on real hardware where photon sources can be probabilistic and lossy. Throughout the paper, we use single-photon input states with $m/2$ photons distributed across alternating modes (i.e.  states of the form $|1,0,1,0,...\rangle$ ). Though we found good results with this input pattern, we also note that other input state configurations likely prove viable. Second, where \cite{BradlerWallner} uses a parity mapping with 4 differently configured optimisation runs to ensure that the entire solution space is reachable, we use trainable bit-flip parameters. The parity mapping scheme requires $m$ photons in $m$ modes to reach the entire solution space. However, trainable bit-flip parameters allow our algorithm to reach the entire solution space with fewer photons\footnote{This can be verified in the trivial case where the probability of flipping any bit is set to 1/2. Any bit string is then equally likely to be reached.}. Moreover, the use of trainable bit-flips allows us to train using only a single optimisation run and a more general class of circuit architectures, instead of the 4 optimisation runs in \cite{BradlerWallner}.

\subsubsection{Training Method}

Here, we provide additional information about the training method used to update the circuit and bit-flip parameters.

We will denote the length-$m$ bit string output of the circuit at any given time by the random variable $X(\bt, \bp)$. Generally, this random variable depends on the architecture of the circuit and the input single-photon state, but we assume these to be fixed at a given iteration. During each update step, we calculate the gradients of the expected cost function value of bit strings generated by the circuit with respect to internal circuit parameters:
$$\frac{d \mathbb{E}[C(X)]}{d\omega} \qquad \omega\in \bp, \bt.$$
We then use gradient descent methods to update each of these parameters according to the gradients calculated.

As in \cite{BradlerWallner} we use a finite difference method to numerically calculate gradients with respect to beamsplitter parameters. For each $\theta$, we apply the following parameter shift rule:
$$\frac{d \mathbb{E}[C(X)]}{d\theta} \approx \frac{1}{\sin(\phi)}\left(\mathbb{E}[C(X_{\theta + \phi})] - \mathbb{E}[C(X_{\theta - \phi})] \right).$$
This is the first-order variant of a general photonic parameter shift rule that can equivalently be used in this setting \cite{giorgio}. The bit-flip probabilities are parametrised with some function $f:\mathbb{R}\rightarrow [0,1]$ in order to encourage desired convergence behaviour. Each bit-flip probability is given through the relation $p_i = f(\alpha_i)$, where $\alpha_i$ are the parameters which we train directly. We will adopt a sigmoid parametrisation $f(\alpha) = \sigma(\alpha) = (1+e^{-\alpha})^{-1}$, encouraging bit-flip convergence towards discrete values. We discuss the motivation for this in section \ref{qual_analysis}. 

The following formula gives us the exact gradient of $E[C(X)]$ with respect to a bit-flip parameter $\alpha$:
$$\frac{d \mathbb{E}[C(X)]}{d\alpha} = \left( \mathbb{E}[C(X_{p = 1})] - \mathbb{E}[C(X_{p = 0})] \right) \frac{df}{d\alpha}.$$
We provide a short derivation of this in Appendix \ref{section:bit_flip_grad_calc}.

Applying the above formulae, to calculate gradients we must estimate values of the form $\left . \mathbb{E}[C(X)] \right|_{\bt, \bp}$. We achieve this by running the circuit in the specified configuration for a pre-specified number of samples, and then calculating the mean cost function value of the samples generated:
$$\frac{1}{S}\sum_{i=1}^SC(X_i) \approx \mathbb{E}[C(X)] \quad \text{ for large }S.$$
As previously mentioned, the cost function values calculated here are also considered towards tracking the optimal solution.

The optimiser we use to update the parameters using these gradients is stochastic gradient descent (SGD) without momentum. We will apply different learning rates for the interferometer parameters and the bit-flip parameters. Our initial small-scale investigation found that SGD outperformed Adam optimisation. However, we leave more comprehensive comparisons to future research.

\subsubsection{Tiling to increase problem sizes}
\label{section:tiling}

\begin{figure}[h]
    \centering
    \includegraphics[width=0.45\textwidth]{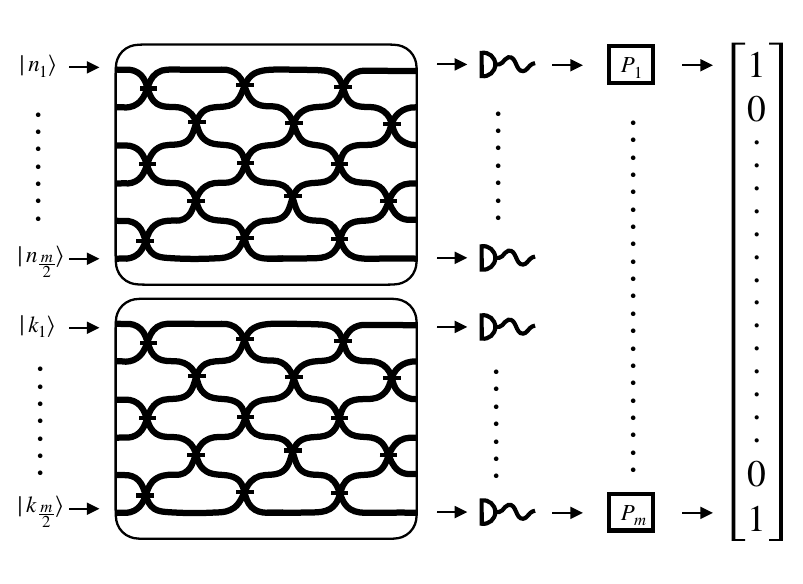}
    \caption{Schematic diagram of a tiled BBS algorithm architecture. To allow a smaller-sized photonic quantum processor to solve a larger problem, we use the quantum processor to implement multiple `tiles', each of which represents a disjoint set of trainable variables.}
    \label{fig:tiling}
\end{figure}

To solve problem sizes that are larger than the actual size of the quantum processor, we propose the use of a `tiling' technique. This involves using several smaller circuits (`tiles') and concatenating their results to produce candidate bit string solutions of the desired scale. This process is illustrated in figure \ref{fig:tiling}. Note that a single physical quantum processor with reprogrammable parameters can implement several tiles. 

For a small number of tiles (up to 3), we found through initial testing that this approach yields good solutions. Though the output distribution from several tiles is separable, since we use a global cost function the gradients of the cost function with respect to each parameter is dependent on all other tiles. This approach is expected to yield lower-quality results than using a larger processor as each tile produces independent samples that are then concatenated together, without correlations between them. We use this method for our experiments with a real quantum processor as a practical way to demonstrate performance on larger problem sizes.

\subsection{Problem formulations}
\label{problem_types_section}
In this section, we introduce the optimisation problems on which we evaluate the BBS algorithm. We select three different problems that are representative of several types of real-world optimisation problems: the knapsack problem, tactical deconfliction, and the travelling salesperson problem.

\subsubsection{The Knapsack Problem}
The knapsack problem is a classical combinatorial optimisation problem with applications in sectors such as investment, logistics, and telecommunications \cite{Knapsack_Applications_Review}. The problem involves choosing a selection of items to maximize their total value while respecting a weight capacity constraint. 

We consider a set of $n$ items, where each item $i$ has value $v_i$  and weight $w_i$. We map a candidate solution to a length $n$ binary string where:
$$
x_i =
\begin{cases}
  1 & \text{if item }i\text{ is included}\\
  0 & \text{otherwise }
\end{cases}
$$
We wish to find the binary string maximising the value:
$$\sum_{i=1}^n x_i v_i,$$
while being subject to the constraint:
$$\sum_{i=1}^n{x_iw_i} \le W.$$
In order to pose this as a binary optimisation task, we define the following cost function acting on the space $\{0,1\}^n$:
$$C(\boldsymbol{x}) = 
\begin{cases}
  \sum_{i=1}^n x_i v_i & \text{if }\sum_{i=1}^n{x_iw_i} \le W\\
  \sum_{i=1}^n x_i v_i - V -1 & \text{otherwise}
\end{cases},$$
where $V =  \sum_{i=1}^n v_i$. This cost function guarantees that all solutions violating the weight constraint return negative cost function values, and valid solutions are costed exactly in accordance with how the problem is posed.

Here, our non-QUBO encoding requires $m = n$ binary variables for an $n$-item problem. We note that QUBO formulations of the discrete knapsack problem require additional variables in the encoding, with the typical QUBO encoding of the problem using $n + \lfloor 1 + \log(W)\rfloor$ binary variables \cite{Ising_Model_NP_Problems}.

We generate instances of the knapsack problem at random by uniformly randomly picking weights and values from a set range.

\subsubsection{Tactical deconfliction}
Tactical aircraft deconfliction is an optimisation problem in which we assign maneuvers to a set of aircraft to ensure that a minimum spatial separation between them is maintained. A violation of this minimal separation is called a conflict. Feasible maneuvers can include speed changes, course angle changes and vertical changes. Previous works on the tactical deconfliction problem \cite{pallottino2002conflict,durand1996automatic,pecyna2024quantum} consider a number of optimisation criteria that include maximizing the number of resolved conflicts with equal impact on flights (fairness), minimizing deviations from nominal aircraft trajectories, and economic criteria such as minimizing fuel consumption or overall delay.
 
We use the same problem formulation as \cite{pecyna2024quantum} where we aim to minimise the number of conflicts between routes, alongside an arbitrary secondary optimisation criterion. We consider a discrete set of $k$ maneuvers for each flight, which includes the option to stay on the original trajectory. For $N$ aircraft, the choice of maneuvers is given by the binary variables:
$$\{ x_{ij} : i = 1,...,N, j = 1, ..., K\},$$
where $ x_{ij}$ is the indicator for whether aircraft $i$ performs maneuver $j$. A valid solution will have exactly one maneuver chosen per aircraft. We thus encode a solution with a length $m = NK$ binary string.

For proposed sets of maneuvers, we calculate a $N \times K \times N \times K$ conflict matrix $CM$ where the entry $CM(i,j,i',j') \in \{0,1\}$ is the indicator function for if aircraft $i$ performing maneuver $j$ conflicts with aircraft $i'$ performing maneuver $j'$.

We construct a cost function consisting of 3 weighted components representing constraints and optimisation goals. The first component checks whether exactly one maneuver is chosen per aircraft. The second component counts the number of conflicts between the chosen trajectories. The final component counts the number of aircraft choosing the `do nothing' maneuver. This is labelled as maneuver 1. This is the ancillary optimisation criterion of minimising change from the original trajectories.

The total cost function for this tactical deconfliction problem is given by:
$$
C(\boldsymbol{x}) = (N \cdot K + 1) \cdot H_1(\boldsymbol{x}) +  (N+1) \cdot H_2(\boldsymbol{x}) - H_3(\boldsymbol{x}),
$$
where we have weighted the terms of the cost function using heuristically chosen coefficients, and the components are defined as:
$$
H_1(\boldsymbol{x}) = \begin{cases}
  0 & \text{if one maneuver chosen per aircraft}\\
  1 & \text{otherwise}
\end{cases}
$$
$$
H_2(\boldsymbol{x}) = \sum_{i=1}^{N} \sum_{j=1}^{K} \sum_{i'=1}^{N} \sum_{j'=1}^{K}C(i,j,i',j')x_{ij}x_{i'j'}
$$
$$
H_3(\boldsymbol{x}) = \sum_{i=1}^N x_{i1}.
$$

We note that this is in fact a QUBO cost function: it can be expressed in the form $\boldsymbol{x}^{T} Q \boldsymbol{x}$ where $Q$ is a symmetric matrix. The exact expression for writing $H_1$ in this form is given in \cite{pecyna2024quantum}.

In the following, we generate random instances of the tactical deconfliction problem by fixing constants $K,N$, and then generating a random symmetric conflict matrix $CM$ where each independent entry is a biased Bernoulli random variable.

\subsubsection{The Travelling Salesperson Problem}

The Travelling Salesperson Problem (TSP) is a common combinatorial optimisation task with applications in areas such as crystallography, scheduling and routing \cite{TSP_Applications_Review}. The problem formulation we consider is the following: given $n$ points on a 2-dimensional plane, find the minimum length path that navigates between all of the points in a cycle. 

We generate an $n$ location instance of the TSP problem by placing $n$ points uniformly at random on a square grid. We generate coordinates $y_i$ for each point.

We encode this problem into binary strings using a permutation encoding. Since every location is visited exactly once, we can arbitrarily fix a point $y_0$ as the start/end location. The rest of the points visited can then be encoded as a permutation of $n-1$ values denoting the order in which they are visited. We generate a permutation $\sigma \in S_{n-1}$. From this, we encode the candidate cycle:
$$y_0 y_{\sigma(1)} ... y_{\sigma(n-1)} y_0.$$
In this way we have a bijection between permutations and candidate solutions to the TSP task. We extend the mapping to binary strings by employing the algorithm introduced in \cite{Binary_to_TSP} which maps the set of binary strings of length $\lceil \log_2((n-1)!) \rceil$ to the set of permutations $S_{n-1}$ surjectively.

We define our cost function in the natural way. Letting $\sigma_{\boldsymbol{x}} \in S_{n-1}$ denote the permutation generated by binary string $\boldsymbol{x}$, we define the cost function as the distance taken to traverse the encoded cycle:
$$C(\boldsymbol{x}) = \sum_{i=0}^{n-1}\Vert y_{\sigma_{\boldsymbol{x}}{(i)}} - y_{\sigma_{\boldsymbol{x}}{(i+1)}}\Vert,$$
where we abuse notation to define the permutation acting outside of its range as $\sigma_{\boldsymbol{x}}(0) ,\sigma_{\boldsymbol{x}}(n) \coloneqq
 0$. We thus encode an $n$-location TSP problem to a binary optimisation task over $m = \lceil \log_2((n-1)!) \rceil$ variables.

This problem formulation requires significantly fewer variables than QUBO encodings of the TSP problem which require $O(n^2)$ binary variables. This quadratic scaling poses challenges for implementations on QUBO-constrained quantum optimisation algorithms \cite{Ising_Model_NP_Problems, TSP_Dwave}. We note that previous work has also considered binary encodings of TSP \cite{QC_for_TSP, Binary_to_TSP}, with some encodings run on an earlier version of the BBS algorithm \cite{Goldsmith}. The non-penalty term encoding presented in \cite{Goldsmith} has the same scaling as the method we outline here.

The scaling of length $m = \lceil \log_2((n-1)!) \rceil$ binary strings restricts the binary problem sizes we can consider for TSP. Where for the other optimisation classes we consider problems of sizes $10,20,30$, for TSP we consider $10,19,29$.

\section{Simulation Experiments}

In the following experiments, we use a simulated quantum optical circuit to validate the structure of the algorithm and evaluate its performance. As previously mentioned, the circuit we simulate consists of a 1-3-9 power-law delay line architecture with $m$ modes and a $|1,0,1,0,...\rangle$ single-photon input state. The same hyperparameters are used for all problems. The main hyperparameters are the number of update steps ($N = 200$), the number of samples used to estimate expectation values ($S = 50$), and the learning rates for both the interferometer parameters (0.01) and bit-flip parameters (0.05). These were selected based on preliminary testing on the knapsack problem.

The metrics we consider are the number of problems for which an optimal solution is found and the average percentage from the optimal solution. For knapsack and TSP problem instances, the average difference from optimal is calculated as the absolute relative error:
$$\Delta = \left\lvert \frac{C_{\text{opt}} - C_{\text{alg}}}{C_{\text{opt}}} \right\rvert,$$
where the optimal solutions $C_{\text{opt}}$ to each problem have been found by a brute force search, and $C_{\text{alg}}$ is the solution found by the algorithm being considered.

For tactical deconfliction instances, due to the minimum cost function value being negative, and the range of cost function values spanning across 0, we redefine absolute relative error to be:
$$\Delta = \left\lvert \frac{C_{\text{alg}} - C_{\text{opt}}}{C_{\text{max}}-C_{\text{opt}}} \right\rvert,$$
where $C_{\text{max}}$ is the maximum value ever taken by the cost function. As such, $\Delta$ represents the distance of the found solution from the optimum with respect to the entire solution space.

Since simulating a quantum optical circuit takes an exponentially increasing amount of time with classical computing, our experiments are limited to problem sizes up to 30 where the best solution can be found with a brute force search. Though size 30 problems are comparatively small, we find that this is a regime that still allows for meaningful benchmarking and comparisons between different algorithms. With our selected hyperparameters, BBS performs a number of cost function evaluations that is orders of magnitude smaller than the problem space (see Appendix \ref{section: num_cost_fn_evals}), such that finding a good solution is still a challenging problem.

\subsection{Qualitative analysis}
\label{qual_analysis}
\begin{figure}[htbp]
    \centering

{\centering \hspace{2em}Loss Evolution\par}
    \begin{subfigure}{\linewidth}
        \centering
        \includegraphics[width=0.9\linewidth]{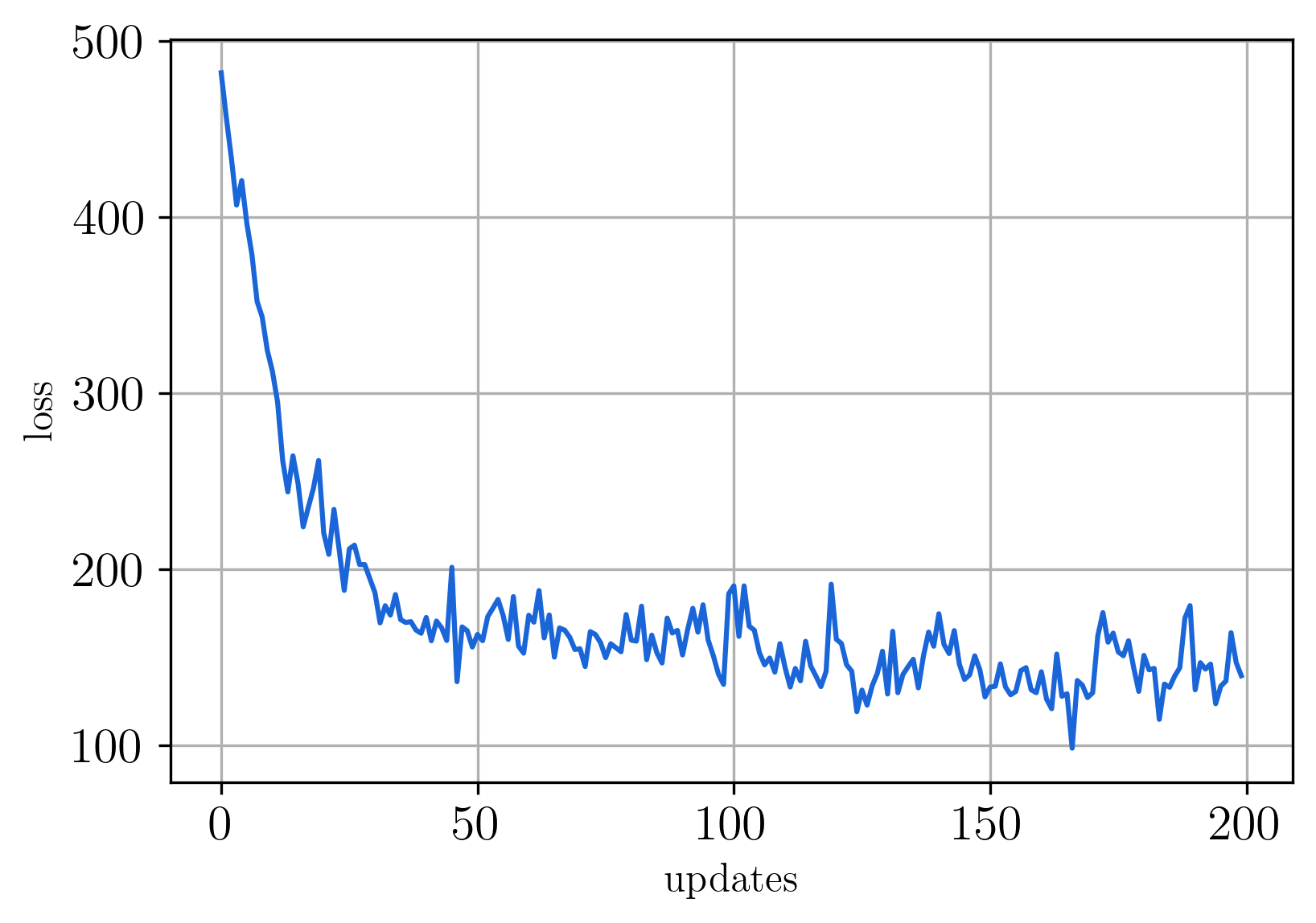}
    \end{subfigure}

    \vspace{0.6em}

{\centering\hspace{2em} Bit-Flip Probabilities\par}
    \begin{subfigure}{\linewidth}
        \centering
        \includegraphics[width=0.9\linewidth]{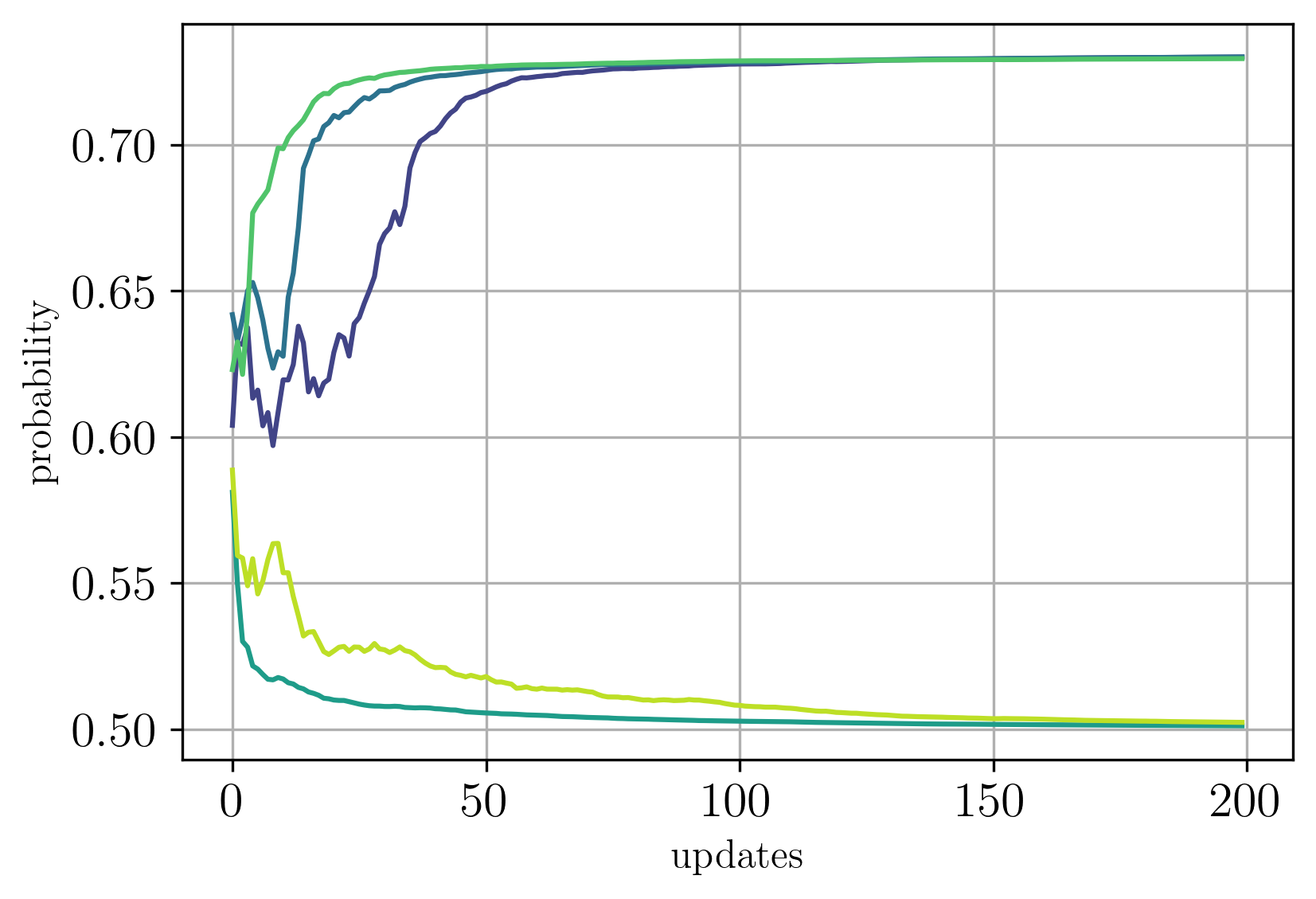}
    \end{subfigure}

    \vspace{0.6em}

{\centering \hspace{2em}Beamsplitter Angles\par}
    \begin{subfigure}{\linewidth}
        \centering
        \includegraphics[width=0.9\linewidth]{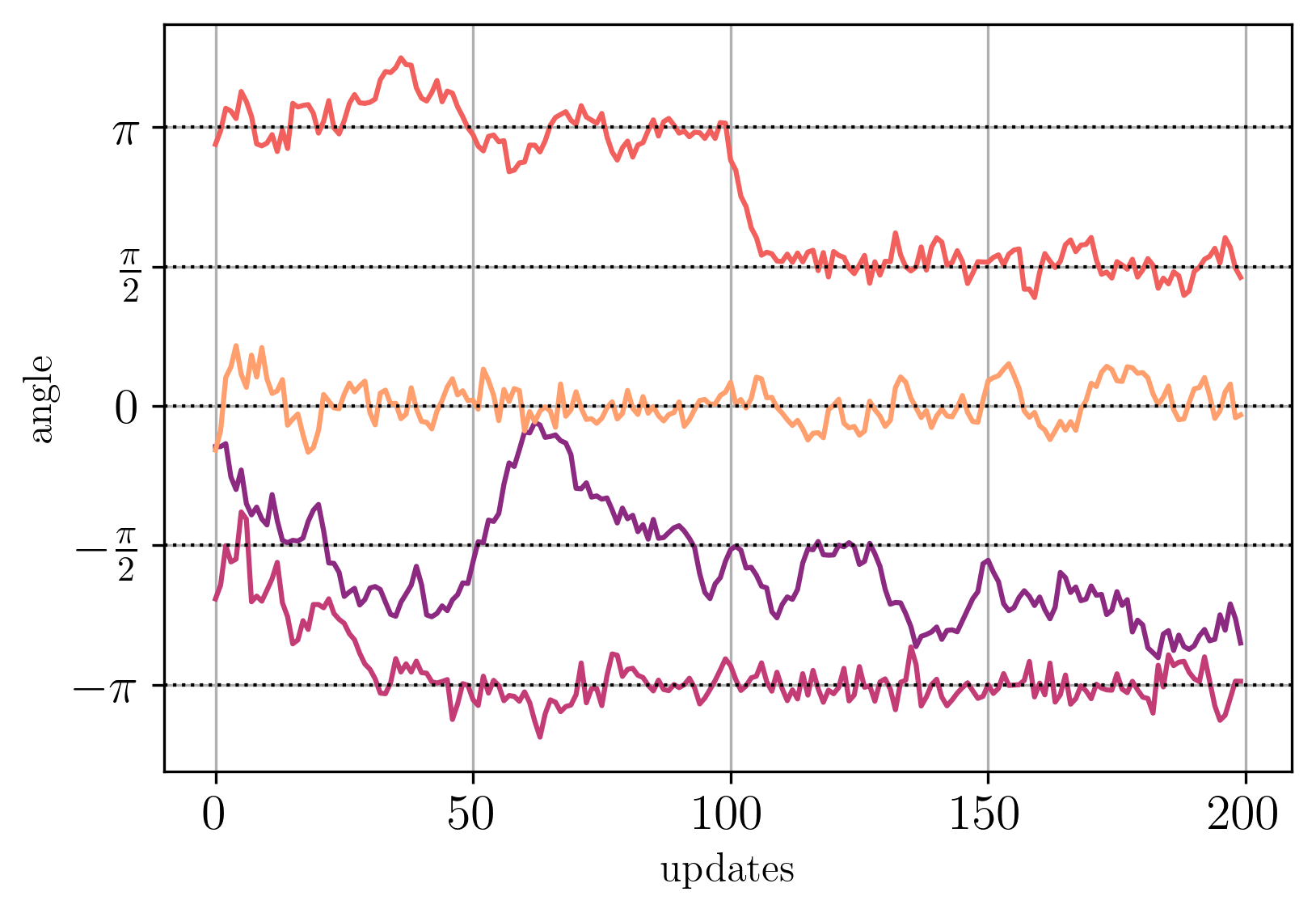}
    \end{subfigure}

    \caption{We plot the evolution of loss and an arbitrary selection of bit-flip probabilities and beamsplitter angles during a typical run of the BBS algorithm. The algorithm is run on a size 30 tactical deconfliction cost function, with the same hyperparameters as previously defined.}
    \label{fig:training_summary}
\end{figure}

In figure \ref{fig:training_summary} we display the evolution of the loss $\mathbb{E}[C(X)]$, and a subset of circuit parameters $\theta_i,p_i$ over the course of a typical BBS optimisation run. Here, we use a size 30 tactical deconfliction cost function, and the hyperparameters are the same as defined at the start of this section.

Over the course of training, we see that the loss value continually decreases. The system initially produces purely random candidate solutions due to the bit-flip probabilities being initialised at $1/2$. With training, the system evolves from producing highly random outputs to focusing on a part of the solution space that contains high-quality (low cost) solutions. Both the bit-flip probabilities and the interferometer parameters contribute to this behavior.

The bit-flip probabilities quickly converge to discrete values corresponding to deterministic flips or no-flips. The convergence to discrete values was intended through the sigmoid parametrisation of these variables. As such, the trained BBS algorithm has learned a deterministic mapping from samples of an appropriately parametrised interferometer to high-quality candidate solutions for the optimisation task at hand. The determinism of the post-processing maintains the sampling complexity of the trained circuit.

In parallel, the algorithm trains to find optimal interferometer parameters. Several parameters converge towards multiples of $\pi/2$, where photons are either deterministically transmitted or reflected to a different mode. As shown by the loss curve, after the bit-flips have been trained to be deterministic, the evolution of the interferometer parameters continues to improve the loss of the circuit.

\subsection{Ablation Testing}
In this section we present ablation tests that validate the design choices in the algorithm. In particular, we highlight the importance of training both the beamsplitter and bit-flip parameters for algorithm performance.

For each experiment, we generate 100 cost functions. Either the full or an ablated BBS algorithm is run on each function and the results compared. Our ablation tests consist of the following:
\begin{itemize}
    \item \emph{non-trained thetas}: learning rates for beamsplitter angles set to 0
    \item \emph{non-trained all}: learning rates for all parameters set to 0
\end{itemize}
We run such ablation tests on each problem type discussed in \ref{problem_types_section}. We note that a \emph{non-trained all} circuit is equivalent to one which generates bit strings uniformly at random as the bit-flips are initialised at $p_i = 1/2$, overriding the statistics of the photonic samples. As such, this ablation test is effectively equivalent to brute force sampling within the solution space with the same maximum number of cost function calls as used by the BBS runs.

\begin{figure}[htbp]
    \centering

{\centering \hspace{2em} Knapsack\par}
    \begin{subfigure}{\linewidth}
        \centering
        \includegraphics[width=0.8\linewidth]{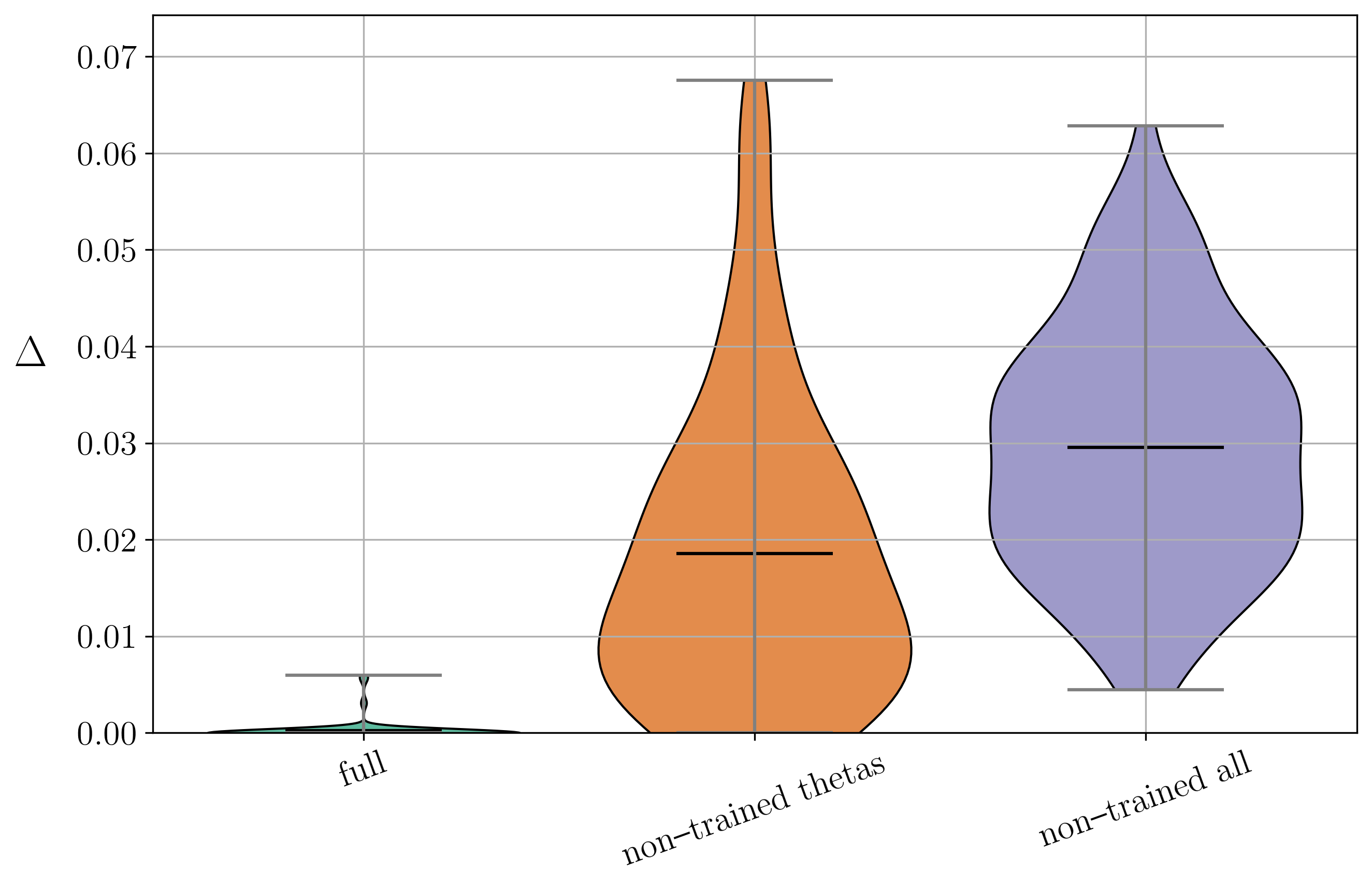}
    \end{subfigure}

    \vspace{0.6em}

{\centering \hspace{2em} Tactical Deconfliction\par}
    \begin{subfigure}{\linewidth}
        \centering
        \includegraphics[width=0.8\linewidth]{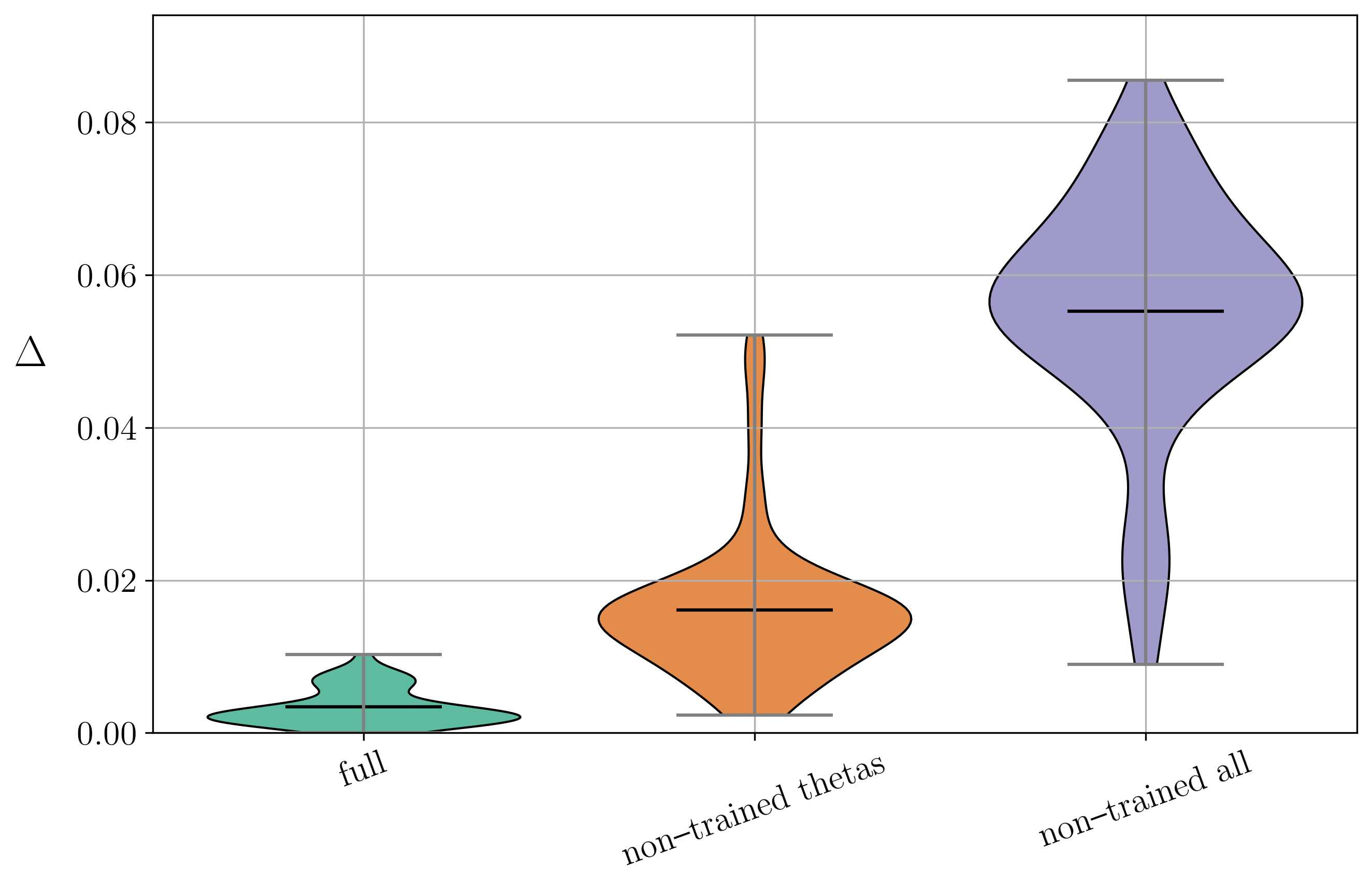}
    \end{subfigure}

    \vspace{0.6em}

    {\centering \hspace{2em}TSP\par}
    \begin{subfigure}{\linewidth}
        \centering
        \includegraphics[width=0.8\linewidth]{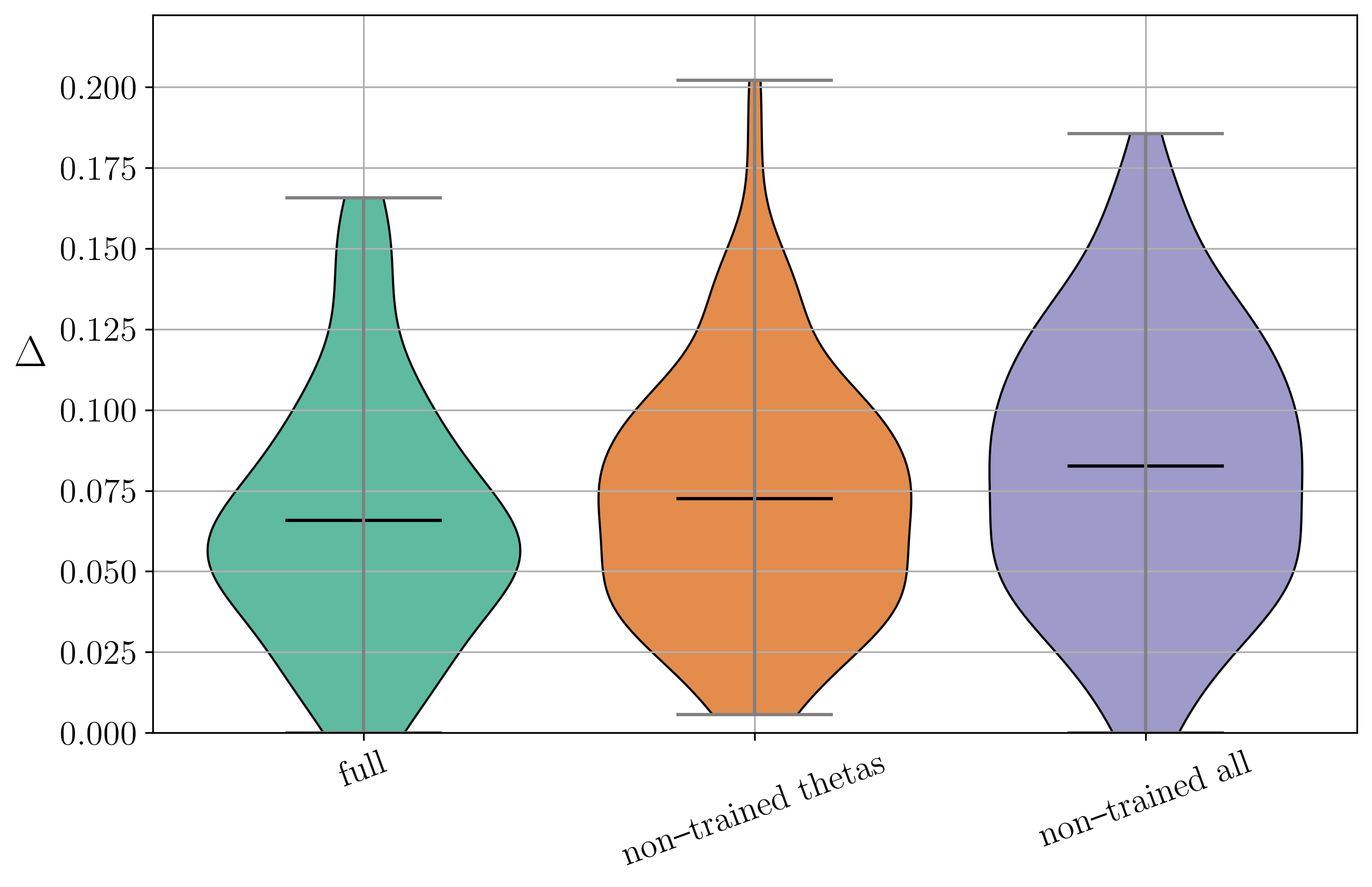}
    \end{subfigure}

    \caption{Violin plots for dimension 30 ablation tests across different optimisation problem types. Each parameter training subroutine of the BBS algorithm improves performance. The magnitude of this improvement is problem dependent.}
    \label{fig:ablation_all}
\end{figure}

Our ablation tests are presented as violin plots in figure \ref{fig:ablation_all}. We observe improvements in average solution quality, variance of solution quality and percentage of optimal solutions when the bit-flip parameters are trained and further again when beamsplitter parameters are trained. The behaviour of the ablation tests varies with the different types of cost function, indicating varying suitability of the algorithm to different problems as can be expected of a heuristic optimisation algorithm. These results validate the underlying design choices for the BBS algorithm.

\subsection{Performance Evaluation}
In this section, we present quantitative experiments examining the quality of solutions found by the full BBS algorithm for increasing problem sizes. We also compare the performance of BBS to two classical heuristic algorithms: simulated annealing (SA) and hill climbing with restarts (HC). As a point of comparison between the algorithms, we give each of them the same maximum number of cost function calls. For BBS, this is calculated as in Appendix \ref{section: num_cost_fn_evals}. Hill Climbing and Simulated Annealing are given this same maximum number of evaluations, as defined for each algorithm respectively. This metric avoids any hardware-dependent considerations.

For each problem type discussed in section \ref{problem_types_section} we generate 100 cost functions and present average percentages from the optimal solution calculated in the same manner as introduced in the start of this section.

\textbf{Simulated Annealing Implementation} We use the standard \emph{simanneal} python module \cite{simanneal} implementing the well-known classical optimisation method of simulated annealing. We use the default hyperparameters and the same cost functions as those fed into the implementation of the Bosonic Binary Solver. We define a `move' in simulated annealing as a flip of a bit in a bit string.

\textbf{Hill climbing with restarts} Our implementation of hill climbing with restarts uses a `first improvement' search approach. We initialise randomly at a bit string. At random, we pick a bit of the string, flip it, and see if the cost function of the new string is lower than the previous one. If so, we switch to the new string. Otherwise, we keep trying the other bits until we exhaust options. Then, we restart at a new random initial bit string. The algorithm is described in appendix \ref{alg:hill_climb}.

\begin{table}[h]
  \centering
  \scriptsize
  \renewcommand{\arraystretch}{1.1}
  \begin{tabular}{|l|c|c|c|c|c|c|}
    \hline
    \makecell{Pb.\\Size} & \makecell{BBS \%\\Optimal} & \makecell{BBS Avg.\\\% Error} & \makecell{SA \%\\Optimal} & \makecell{SA Avg.\\\% Error} & \makecell{HC \%\\ Optimal} & \makecell{HC Avg.\\\% Error} \\
    \hline
    \multicolumn{7}{|c|}{\textbf{Knapsack}} \\
    \hline
    10 & 100 & 0.00 & 100 & 0.00 & 100 & 0.00 \\
    15 & 100 & 0.00 & 100 & 0.00 & 100 & 0.00 \\
    20 &  98 & 0.01 & 100 & 0.00 & 84 & 0.23 \\
    25 &  99 & 0.00 & 100 & 0.00 & 13 & 2.88 \\
    30 &  93 & 0.06 & 100 & 0.00 & 23 & 1.64 \\
    \hline
    \multicolumn{7}{|c|}{\textbf{Tactical Deconfliction}} \\
    \hline
    10 & 100 & 0.00 & 100 & 0.00 & 100 & 0.00 \\
    15 & 100 & 0.00 & 100 & 0.00 & 100 & 0.00 \\
    20 &  86 & 0.07 & 100 & 0.00 & 100 & 0.00 \\
    25 &  78 & 0.08 &  99 & 0.01 & 100 & 0.00 \\
    30 &  33 & 0.13 &  92 & 0.01 & 100 & 0.00 \\
    \hline
    \multicolumn{7}{|c|}{\textbf{TSP}} \\
    \hline
    10 & 100 & 0.00 & 100 & 0.00 & 100 & 0.00 \\
    19 &  99 & 0.01 & 97 & 0.07 & 100 & 0.00 \\
    29 &   2 & 6.58 &  27 & 2.94 & 41 & 1.36 \\
    \hline
  \end{tabular}
  \caption{Comparison of the performance of BBS, Simulated Annealing and Hill Climbing across problems. All three algorithms manage to find some optimal solutions for the largest sized problem instances. Comparative performance between algorithms differs by problem set.}
  \label{tab:comparison_all}
\end{table}

Our results are shown in table \ref{tab:comparison_all}. When we quote average \% error, this is equal to $100\cdot \Delta$ for the average distances from optimal defined previously.

First, we find that the BBS algorithm is able to find at least some optimal solutions for all problem sizes. For problem sizes up to 25, more than half of BBS runs return an optimal solution. For larger instance sizes, the efficiency of the algorithm in finding the optimal solution decreases. However, the average relative error of the solution found by the algorithm is still low, especially for knapsack and tactical deconfliction, which indicates that the solutions found are close to the optimum. This is a promising result for a heuristic algorithm where finding a good but not optimal result is often acceptable.

Comparing the effectiveness of the BBS algorithm to simulated annealing and hill climbing, we find heterogeneous results that reflect the diversity of the problem set. BBS is found to outperform hill climbing on the knapsack problem, and underperform simulated annealing. Hill climbing is found to work best for both tactical deconfliction and TSP. Though BBS generally underperformed SA, interestingly we found that the problem instances on which BBS found an optimal solution are often different to those on which SA found an optimal solution (see Appendix \ref{SAvsBBS}). These results show that BBS is a viable algorithm for discrete optimisation and could be a useful addition to a heuristic optimisation toolkit. As discussed in the conclusion, we also anticipate that algorithmic improvements could also further improve the performance of BBS.

\section{Hardware Experiments}

In this section we demonstrate a similar experiment running on a real photonic quantum processor. We used two ORCA PT-1 photonic quantum computing devices installed at Poznańskie Centrum Superkomputerowo-Sieciowe (Poznan Supercomputing and Networking Center, PCSS). Both machines can run experiments with up to 4 photons in 8 modes in a time-bin configuration with up to two sequential delay lines of the same length. For our experiments we used only one of the two loops. This trades off decreased programmability and interference range due to fewer beamsplitter parameters for faster runtimes due to less system loss. We further change the number of update steps to $N = 50$, and the number of samples for estimating expectation values to $S = 20$, to reduce the time required to run the experiments. This means that the number of cost functions being evaluated during the hardware experiments is significantly lower than what was used for the simulations in the previous section (see Appendix \ref{section: num_cost_fn_evals}). This will reduce performance compared to an equivalent simulated run. For instance sizes larger than $8$ we used the ‘tiling’ technique described in section \ref{section:tiling} with up to 4 tiles. For each experiment, we generated 10 different cost functions.

\begin{table}[h!]
  \centering
  \begin{tabular}{|l|c|c|}
    \hline
    Pb. Size & BBS \% Optimal & BBS Avg. \% Error \\
    \hline
    \multicolumn{3}{|c|}{\textbf{Knapsack optimisation}} \\
    \hline
    6  & 100 & 0.00 \\
    12 & 100 & 0.00 \\
    18 &  80 & 0.44 \\
    \hline
    \multicolumn{3}{|c|}{\textbf{Tactical Deconfliction}} \\
    \hline
    6  & 100 & 0.00 \\
    12 & 100 & 0.00 \\
    18 &  50 & 0.30 \\
    \hline
    \multicolumn{3}{|c|}{\textbf{TSP}} \\
    \hline
    5  & 100 & 0.00 \\
    10 & 100 & 0.00 \\
    16 &  60 & 3.89 \\
    \hline
  \end{tabular}
  \caption{BBS performance on ORCA PT-1 systems across different problems. Hardware runs of the algorithm successfully find optimal solutions for problems of up to 18 variables in size.}
  \label{tab:formatted_bbs_table}
\end{table}

We find that the algorithm running on real hardware successfully solves problem sizes with up to 18 variables despite experimental imperfections and the use of tiling. This validates the use of this algorithm on real hardware, and the potential of the tiling method to solve problem sizes larger than the size of the quantum processor.

\section{Discussion}

Our work shows that the BBS algorithm is a promising approach for leveraging photonic quantum processors to solve discrete optimisation problems. Since photonic quantum processors have been shown to scale to large sizes, the BBS algorithm is uniquely well suited for scaling to large-scale problems on real hardware. Moreover, since the types of problem that can be solved are not restricted to specific formulations such as QUBO, BBS also sidesteps problem encoding overheads that can affect other types of quantum optimisation algorithms. 

An important consideration for a heuristic optimisation algorithm is the problem type for which it is best suited. In our work, we focused on the setting in which cost functions are easy to evaluate but the problem space is exponentially large. We also focused on problem formulations that use as few binary variables as possible, which helps address larger problem sizes. We expect that efficient problem encodings will remain an important factor in determining the performance and scalability of this algorithm.

Evaluation methods are also an important consideration. Here, we focus on solution quality for a given number of cost function evaluations. We do not consider other metrics such as wall clock time, since our simulations are limited by the computational complexity of simulating a quantum system, and our experiments on real hardware are limited by the specifications of the PT-1, which is a first-generation quantum system. The wall clock time is likely to vary significantly between different hardware systems, while the number of function evaluations used in our work is independent of hardware specifications.

\vspace{0.5cm}
\section{Conclusion}

In this work, we proposed a new Bosonic Binary Solver hybrid quantum-classical algorithm for discrete optimisation that leverages the capabilities of near-term photonic quantum processors. We find in simulation and on real hardware that this algorithm finds good solutions to three different families of optimisation problems. Leveraging the scalable nature of photonics as a quantum computing platform, we anticipate that this algorithm could be applied to large problems in the near term.

We note that there are several potential avenues for further research. For example, developing a better theoretical understanding of when sampling from a non-classical distribution can lead to improved optimisation results would be beneficial. Several options also exist for algorithmic improvements. First, our general variational optimisation scheme is flexible and could be used with a wide range of input states and classical post-processing techniques. Second, several alternatives exist to gradient-based optimisation of a quantum processor, and other non-gradient based techniques could be advantageous. Moreover, developing problem formulations that are better suited for BBS could also lead to improved results. We leave these questions to future research.

\begin{acknowledgments}

We thank Alex Jones and Omar Bacarreza for insightful discussions and comments. The authors also thank Konrad Wojciechowski, Tomasz Pecyna, and Grzegorz Waligóra for their valuable support with reference experiments and the development of benchmarking methodologies for hybrid quantum–classical systems setup at PCSS. This research was funded by the Collaboration Agreement no. 323/PCSS/2024 between ORCA and PCSS, and supported by the Polish Ministry of Science and Higher Education through the “Applied Doctorate” program (agreement no. DWD/6/0142/2022), carried out at the Poznan University of Technology (project no. 0311/SBAD/0746). All computational experiments were conducted on the PCSS quantum–classical setup as part of the EuroHPC PL national infrastructure in Poland. 

\end{acknowledgments}
\bibliographystyle{ieeetr}
\bibliography{biblio.bib}

\clearpage

\appendix
% \begin{center}
% 	{\Large \bf Appendix} 
% \end{center}

\renewcommand{\thesubsection}{\thesection.\alph{subsection}}

\section{Calculating Gradients for Bit-Flips}
\label{section:bit_flip_grad_calc}
Here, we provide a short derivation of the formula used to estimate the gradient of the expected cost function value of the BBS circuit with respect to a bit-flip probability parametrising variable $\alpha_i$.
$$E[C(X)]\big|_{p_i = x} = x\cdot E[C(X)]\big|_{p_i = 1} + (1-x)\cdot E[C(X)]\big|_{p_i=0}$$
$$\Rightarrow \frac{dE[C(X)]}{dp_i} = E[C(X)]\big|_{p_i = 1} - E[C(X)]\big|_{p_i=0}.$$
Consider the scenario where the probabilities are parametrised through the relation $f(\alpha_i) = p_i$. Substituting in this relation we derive:
$$\frac{dE[C(X)]}{d\alpha_i} = f'(\alpha_i)\left(E[C(X)]\big|_{p_i = 1} - E[C(X)]\big|_{p_i=0}\right).$$
\section{Bounding Cost Function Evaluations}
\label{section: num_cost_fn_evals}
We present a formula for upper bounding the number of cost function evaluations used by a run of the BBS algorithm. This is an upper bound since, in practice during a run of the BBS algorithm, the same bit string may be generated several times over as a candidate solution and the corresponding cost function value retrieved from a cache.

Consider running the BBS algorithm with the following parameters:
\begin{itemize}[noitemsep, topsep=0pt]
    \item Number of updates: $N$
    \item Number of samples: $S$
    \item Number of modes (i.e. problem size) : $m$
    \item Number of loops : $K$
    \item Loop lengths : $\{ l_1, ..., l_K\}$
\end{itemize}
The upper bound for the number of cost function evaluations is given by:
$$NS \left( 2\left[\sum_{i=1}^K(m-l_i) \right] + 2m + 1\right).$$

During an update in the BBS algorithm, we need to calculate gradients of the expected cost function value with respect to every single beamsplitter parameter and bit-flip probability. We then run
the whole algorithm as a \emph{forward pass} to get the average cost function value of the
samples generated by the circuit after that update step.

We estimate the expectation values appearing in these calculations by averaging cost function values over $S$ samples from an appropriately parametrised circuit:
$$\frac{1}{S}\sum_{i=1}^SC(X_i) \approx \mathbb{E}[C(X)] \quad \text{ for large }S.$$

\textbf{Cost Function Calls for Beamsplitter Updates}:
We recall that we estimate gradients with respect to beamsplitter parameters with a first-order photonic parameter shift rule \cite{giorgio}:
$$\frac{d \mathbb{E}[C(X)]}{d\theta} = \frac{1}{\sin(\phi)}\left(\mathbb{E}[C(X_{\theta + \phi})] - \mathbb{E}[C(X_{\theta - \phi})] \right).$$

According to the structure of time-bin interferometers, the total number of beamsplitters in the circuit is equal to:
$$\sum_{i=1}^K (m-l_i).$$
We estimate the gradient formula for all of these parameters during each update step. Thus, in total over the course of the algorithm, the number of cost function evaluations used for these estimates is given by:
$$2NS\sum_{i=1}^K (m-l_i).$$
\textbf{Cost Function Calls for Bit-Flip Probability Updates}:
Recall the gradient formula used for bit-flips, similarly with two cost function expectation values present:
$$\frac{dE[C(X)]}{dp_i} = E[C(X)]\big|_{p_i = 1} - E[C(X)]\big|_{p_i=0}.$$
There are $m$ total bit-flip parameters (one per binary variable) and so over the course of the algorithm the maximum number of cost function calls used for bit-flip gradient
calculation is given by:
$$2NSm.$$
\textbf{Cost Function Calls for Forward Passes}:
At the end of each update step, we estimate $$\mathbb{E}[C(X)],$$ as a measure of performance. As such, over the course of the whole algorithm the
forward pass uses at most $NS$ cost function calls.

We add all of these values together to get the initially presented bound. 

This upper bound allows us to quickly bound the proportion of the solution space a run of the BBS algorithm can ever explore. Consider the exact setup used within the performance benchmarks presented in \ref{tab:comparison_all}: we have three delay lines of lengths $\{1,3,9\}$, $N = 200$, $S = 50$.

For a size 30 problem ($m = 30)$, using the above formula we can strictly bound the number of possible candidate solutions by $2.15 \cdot 10^6$.
The number of possible solutions is equal to $2^{30}$. Thus, we explore a maximum of $\approx 0.2\%$ of the entire solution space.

\section{Comparing solutions found by Simulated Annealing and BBS}
\label{SAvsBBS}

\begin{table}[h]
  \centering
  \scriptsize
  \begin{tabular}{|l|c|c|c|}
    \hline
    Pb. Size & \makecell{BBS \%\\Optimal} & \makecell{SA \%\\Optimal} & \makecell{Combined\\\% Optimal} \\
    \hline
    \multicolumn{4}{|c|}{\textbf{Knapsack}} \\
    \hline
    10 & 100 & 100 & 100 \\
    15 & 100 & 100 & 100 \\
    20 &  98 & 100 & 100 \\
    25 &  99 & 100 & 100 \\
    30 &  93 & 100 & 100 \\
    \hline
    \multicolumn{4}{|c|}{\textbf{Tactical Deconfliction}} \\
    \hline
    10 & 100 & 100 & 100 \\
    15 & 100 & 100 & 100 \\
    20 &  86 & 100 & 100 \\
    25 &  78 &  99 & 100 \\
    30 &  33 &  92 & 97 \\
    \hline
    \multicolumn{4}{|c|}{\textbf{TSP}} \\
    \hline
    10 & 100 & 100 & 100 \\
    19 & 99 & 97  & 100 \\
    29 & 2  & 27  & 29 \\
    \hline
  \end{tabular}
  \caption{Comparison of the overlap in optimal solutions found by BBS and Simulated Annealing across problems. Every problem class that is not completely solved by both algorithms exhibits problem instances solved by one, but not the other.}
  \label{tab:combined_optimal}
\end{table}

Here, we investigate the extent to which the problem instances solved by SA and BBS are the same or different. This is to distinguish between two possible explanations for the observation that SA finds optimal solutions more often that BBS. If the problems optimally solved by BBS are also solved by SA, this would suggest that some problem instances are easier and can be solved by both algorithms, whereas some harder problems can only be solved by SA. This would imply that SA is a strictly better general-purpose algorithm. However, if the problems solved by both algorithms are different, then there are cases where BBS finds a better solution than SA.

Table \ref{tab:combined_optimal} shows the number of problems for which BBS, SA, and both algorithms combined found the optimal solution. We find that the number of optimal solutions found by the combined algorithms is larger than the number of optimal solutions found only by SA. This shows that BBS finds optimal solutions to problems where SA does not find an optimal solution. As such, it could be beneficial to use both algorithms side-by-side within a heuristic optimisation toolkit.

\vspace{0.5cm}
\section{Hill Climbing Algorithm}
\label{alg:hill_climb}

\begin{algorithm}[H]
\caption{Hill Climbing Implementation}
\begin{algorithmic}[1]
\Statex \hspace*{-\algorithmicindent} \textbf{Inputs:} Cost function $C$, bit string length $m$, maximum number of cost function evaluations $R$
\Statex \hspace*{-\algorithmicindent} \textbf{Outputs:} Lowest cost value found and corresponding solution $C_{\text{alg}}, \mathbf{x}_{\text{alg}}$
\State{$C_{\text{alg}} \leftarrow \infty$}
\State{$ r \leftarrow 0$}\Comment{Current number of cost functions evaluated}
\While{$r < R$}
    \State $\mathbf{x} \leftarrow$ uniformly random bit string of length $m$
    \State $C_{\text{curr}} \leftarrow C(\mathbf{x})$
    \State $r \leftarrow r + 1$
    \State $\mathcal{T} \leftarrow \{1,...,m\}$ \Comment{Bits not tried in current iteration}
    \While{$\mathcal{T} \ne \emptyset$}
        \State Select $i \in \mathcal{T}$ uniformly at random
        \State $\mathbf{x}' \leftarrow \mathbf{x}$ with $i$th bit-flipped
        \State $C' \leftarrow C(\mathbf{x}')$
        \State $r \leftarrow r + 1$
        \If{$C' < C_{\text{curr}}$}
            \State $\mathbf{x} \leftarrow \mathbf{x}'$
            \State $C_{\text{curr}} \leftarrow C'$
            \State $\mathcal{T} \leftarrow \emptyset$ 
        \Else
            \State $\mathcal{T} \leftarrow T-i$
        \EndIf
    \EndWhile
    \If{$C_{\text{curr}} < C_{\text{alg}}$}
        \State $C_{\text{alg}} \leftarrow C_{\text{curr}}$
        \State $\mathbf{x}_{\text{alg}} \leftarrow \mathbf{x}$
    \EndIf
\EndWhile
\Statex \hspace*{-\algorithmicindent} \textbf{Return} $\mathbf{x}_{\text{alg}}, C_{\text{alg}}$
\end{algorithmic}
\end{algorithm}

\end{document}